\documentclass[runningheads]{llncs}

\usepackage{subfig}
\usepackage{caption}
\usepackage{graphicx}
\newcommand{\fix}[2]{{\bf FIX}\footnote{{\bf #1:} #2 }}
\renewcommand{\fix}[2]{}

\begin{document}
\title{Explaining Cybersecurity \\ with Films and the Arts \\ {\normalsize (Extended Abstract)}\thanks{This paper was presented at the Imagine Math 7
Conference ``Mathematics and Culture'', Venice, March 29--31, 2019.}}
%
\author{Luca Vigan\`o}
\authorrunning{Luca Vigan\`o}
%
\institute{Department of Informatics \\ 
King's College London \\ 
London, UK \\
luca.vigano@kcl.ac.uk}

\maketitle              


\section{Introduction}
There are a large number of movies, TV series, novels and even plays about cybersecurity and, in particular, about hackers. Some are good, some are so-so, most are frankly quite bad. Some are realistic, most make cybersecurity experts cringe. 

Inspired by DARPA's \emph{Explainable Artificial Intelligence (XAI)} program, Da\-niele Magazzeni and I proposed a new paradigm in security research: \emph{Explainable Security (XSec)}. In~\cite{XSec}, we discussed the ``Six Ws'' of XSec (Who? What? Where? When? Why? and How?) and argued that XSec has unique and complex characteristics: XSec involves several different stakeholders (i.e., the system's developers, analysts, users and attackers) and is multi-faceted by nature (as it requires reasoning about system model, threat model and properties of security, privacy and trust as well as concrete attacks, vulnerabilities and countermeasures). We defined a roadmap for XSec that identifies several possible research directions. 

In this paper, I address one of these directions.
I discuss how several basic cybersecurity notions (and even some advanced ones) can be explained with the help of some famous and some perhaps less obvious popular movies and other artworks.
I will focus in particular on anonymity, pseudonymity and authentication, but similar explanations can be given for other security properties, for the algorithms, protocols and systems that have been developed to achieve such properties, and for the vulnerabilities and attacks that they suffer from.

In~\cite{XSec}, we gave a detailed list of who gives the explanations but also who receives them, and pointed out that the recipients of the explanations might be varied, ranging from experts to laypersons; in particular, non-expert users will need to receive an explanation of how to interact with a security system, why the system is secure and why it carries out a particular action. To that end, it will be necessary to explain the security notions, properties and mechanisms in a language, and in a way, that is understandable by the laypersons. In practice, however, users are rarely given such explanations and thus end up being frustrated and disillusioned by security systems, which might lead them to committing mistakes that make the systems vulnerable to attacks.
Clear and simple explanations with popular movies and the arts allow experts to target the laypersons, reducing the mental and temporal effort required of the laypersons and increasing their understanding, and ultimately their willingness to engage with security systems.

\section{Security properties explained}

\begin{figure}[t]
\centering
\includegraphics[width=0.68\textwidth]{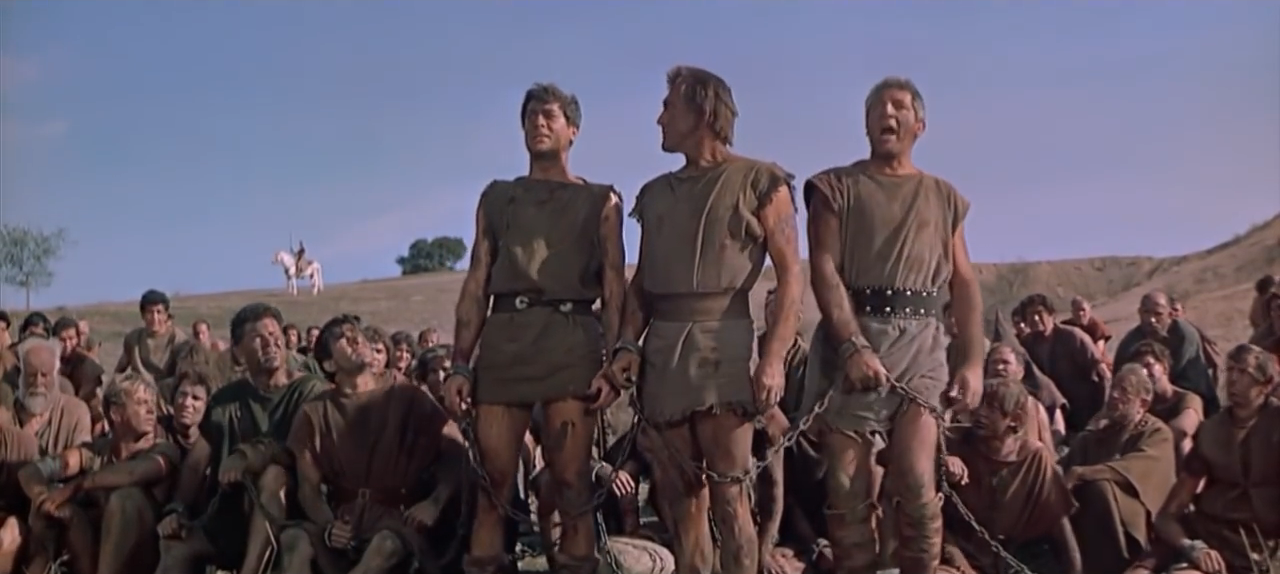}
\caption{I'm Spartacus!}
\label{fig:IamSpartacus}
\end{figure}

Picture this. A battlefield, 
71 B.C. The Roman soldiers have crushed the revolt of the gladiators and slaves led by Spartacus. Many of the rebels have been killed in the battle, while Spartacus, his right-hand man Antoninus and the other surviving rebels have been captured and are sitting in chains on a hillside. There is just one catch: the Romans don't know who Spartacus is. So, the Roman general addresses the prisoners: 
\begin{quote}\em
I bring a message from your master, Marcus Licinius Crassus, commander of Italy. By command of his most merciful excellency, your lives are to be spared. Slaves you were and slaves you'll remain, but the terrible penalty of crucifixion has been set aside under single condition that you identify the body or the living person of the slave called Spartacus. 
\end{quote}
A beat.
Spartacus stands up and opens his mouth to speak, but Antoninus leaps to his feet and shouts ``I'm Spartacus!'', quickly followed by the slave to Spartacus' left (see Fig.~\ref{fig:IamSpartacus}). To the bewilderment of the real Spartacus and of the Roman general, soon everyone around Spartacus is shouting ``I'm Spartacus!'' (well, actually, everyone except Spartacus himself). 

\begin{figure}[!t]
\centering
\includegraphics[width=0.685\textwidth]{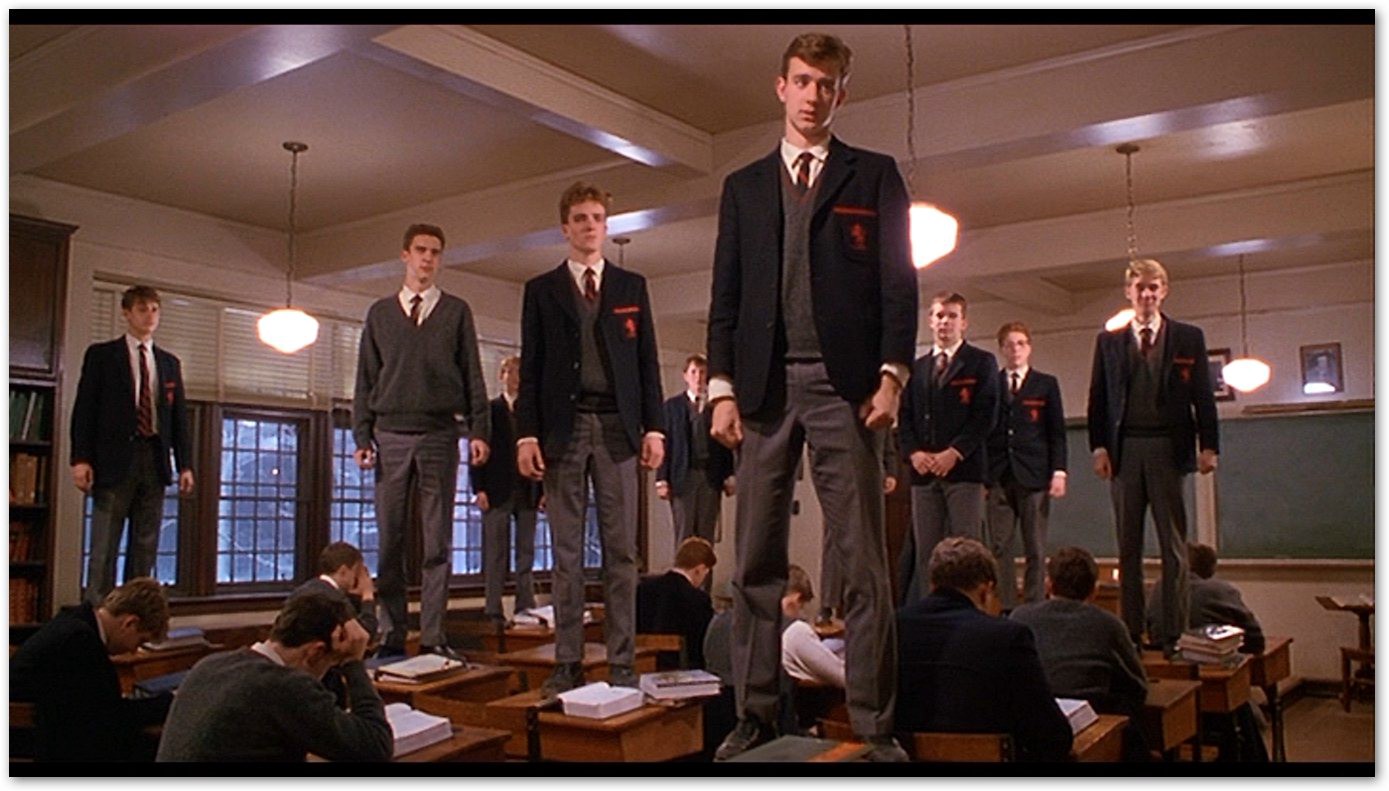}
\caption{O Captain! My Captain!}
\label{fig:DeadPoetsSociety}
\end{figure}

This is probably the most famous scene of the movie \emph{Spartacus}~\cite{Spartacus}, which has been imitated in several other movies such as in the ``O Captain! My Captain!'' scene in \emph{Dead Poets Society} (Fig.~\ref{fig:DeadPoetsSociety}, \cite{DeadPoetsSociety}).\footnote{The
original novel by Howard Fast contains a similar narration, but the script by Dalton Trumbo and the mise-en-sc\`ene by Stanley Kubrick turned it into a crisper and more powerful scene.}

What just happened? In order to protect Spartacus, Antoninus and the other prisoners have created a what in technical terms is called an \emph{anonymity set}: given that all the prisoners are claiming to be Spartacus, it is impossible for the Romans to identify who is the real Spartacus. In other words, Spartacus is \emph{anonymous} as he is not identifiable within the set of the prisoners.

\begin{figure}[!t]
\centering
\includegraphics[width=0.4\textwidth]{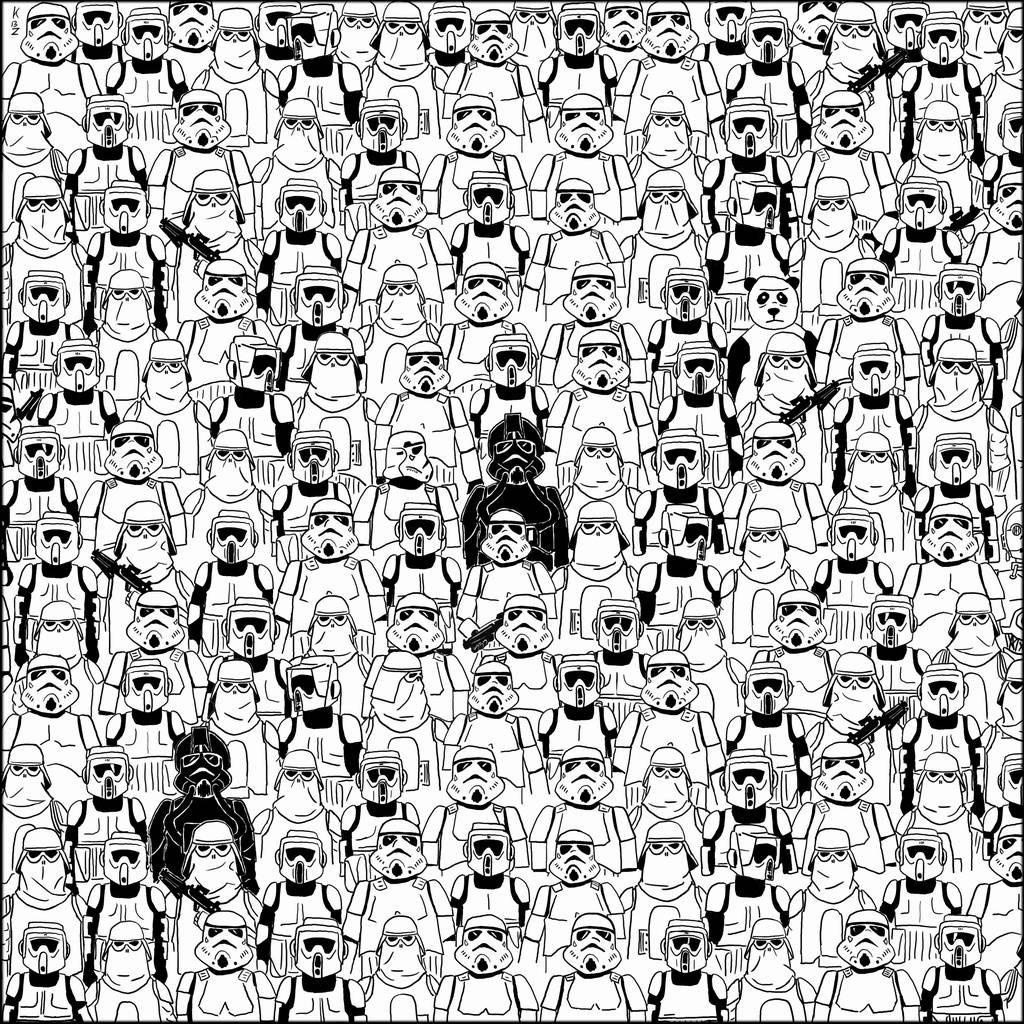}
\caption{Find the Panda}
\label{fig:panda}
\end{figure}

One can of course give a formal definition of 
anonymity~\cite{PfitzmannHansen2010} 
--- and this is what one would do in front of an expert audience, say in a talk at a conference or in a university lecture on cybersecurity -- but the point here is that even 
non-specialists immediately understand what Antoninus and the other prisoners are doing and why. 
In fact, I dare say that we all more or less intuitively understand that anonymity cannot exist in a vacuum (one cannot be anonymous by oneself) but rather requires a large enough set of similar ``things'', a large enough set of similar people, actions, messages, etc. so that one's identity, actions or messages are not distinguishable from those of the others and thus not identifiable. This intuition is so common that we have even invented games and puzzles that exploit anonymity sets such as the \emph{Find the Panda} puzzle in Fig.~\ref{fig:panda} (created by reader ``ste1'' of the ``Bored Panda'' website, \url{https://www.boredpanda.com/find-the-panda-illustrated-puzzle-star-wars-edition-ste1/}).\fix{Luca}{Talk about difference with respect to Where's Waldo, where Waldo is hiding among different figures, whereas here the panda is among similar figures.}

\begin{figure}[!t]
\centering
    \subfloat[Thomas Crown]
    {
    \includegraphics[width=0.212\textwidth]{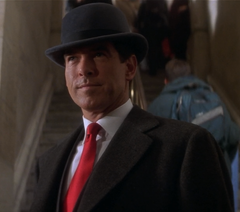}
    \label{fig:ThomasCrown}
    }
    \quad
    \subfloat[The Bowler Hat Men]
    {
    \includegraphics[width=0.37\textwidth]{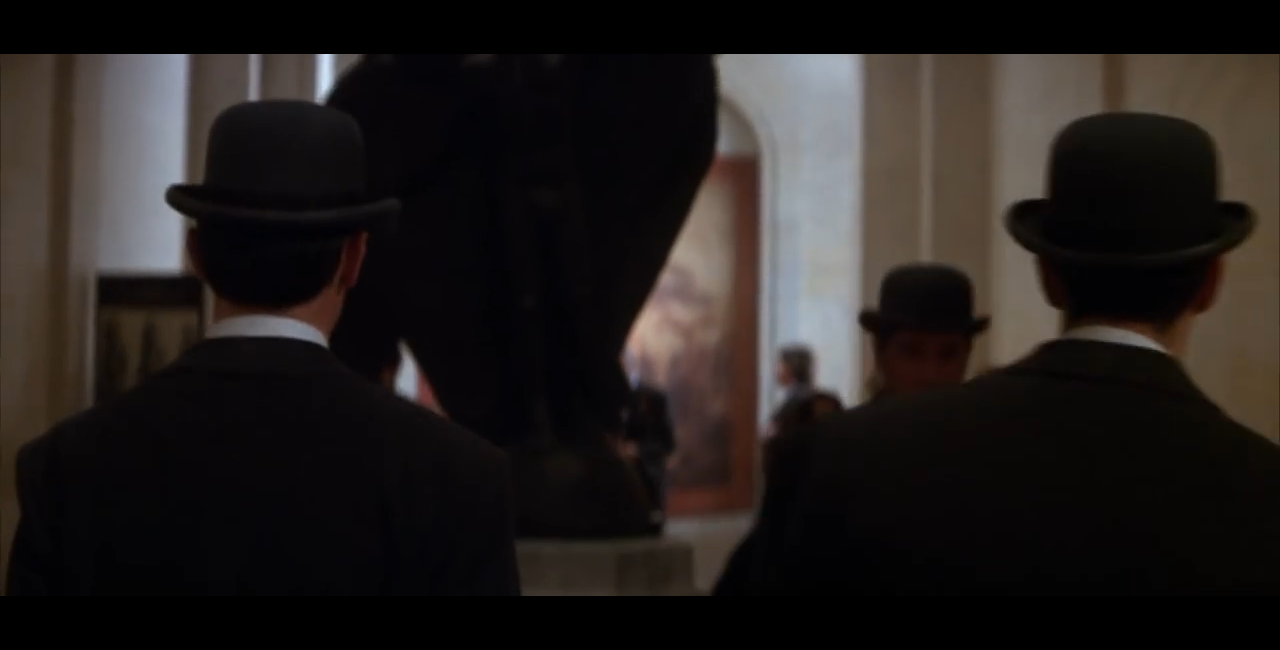}
    \label{fig:ThomasCrown-hats}
    }
    \quad
    \subfloat[The Son of Man]
    {
        \includegraphics[width=0.235\textwidth]{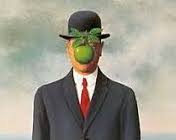}
        \label{fig:TheSonOfMan}
    }
\caption{Unobservability in ``The Thomas Crown Affair''}
\label{fig:TC}
\end{figure}

As another example, consider the climax scene 
\emph{}at the end of \emph{The Thomas Crown Affair}~\cite{ThomasCrown1999}, in which Thomas Crown returns a painting that he has stolen at the beginning of the movie from the Metropolitan Museum of Art in New York City. Even though he had announced that he would return the painting, Thomas Crown manages not to get caught by being \emph{unobservable}: he has hired a number of ``extras'' to walk around in the museum carrying the same briefcase that he does and being dressed exactly him, namely, like the man in a bowler hat in Ren\'e Magritte's \emph{The Son of Man} (see Fig.~\ref{fig:TC}). This is (more or less exactly) the way in which \emph{Mix Networks} realize unobervability and untraceability of messages~\cite{Chaum81}, and thus provide a basis for anonymous communication, emailing and voting.\footnote{Interestingly, that scene of \emph{The Thomas Crown Affair}~\cite{ThomasCrown1999} also provides a very intuitive example for \emph{steganography}, which is the practice of concealing a message, file, image, or video within another message, file, image, or video.}

Similar intuitive definitions can be given of other security properties. For instance, since anonymity is difficult to achieve (as it requires building a very large anonymity set) and is often fraught with legal issues, one could be satisfied with being \emph{pseudonymous}, which is a state of disguised identity (whereas anonymity may be defined as a state of non-identifiable, unreachable, untrackable, or unlinkable identity)~\cite{PfitzmannHansen2010}. The use of \emph{pseudonyms} and the difficulty of carrying out \emph{pseudonym resolution} to reconcile a pseudonym to its true name\footnote{Similarly, \emph{name resolution} is a method of reconciling an IP address to a user-friendly computer name.} can also be explained by means of films and the arts. For example, one could ask the question ``Who is Batman?''. The answer could be ``Bruce Wayne'', referring to the wealthy industrialist who dons the batsuit to fight crime, or it could be ``Adam West, Michael Keaton, Val Kilmer, George Clooney, Christian Bale or Ben Affleck'', the main actors 
who played Batman in TV shows and movies (I am not including Lego Batman, voiced by Will Arnett, although perhaps I should). Both are correct answers, depending on the context and on time, much in the same way pseudonym resolution and name resolution work on the Internet.\footnote{This is tightly connected to the \emph{metaphysics of identity}, as exemplified for instance by the paradox of the \emph{Ship of Theseus}, which is commonly attributed to the Greek philosopher Plutarch.
}

One could similarly ask ``Who is Superman?'', and receive as an answer the names of the actors who played Superman (``Kirk Alyn, George Reeves, Christopher Reeve, Brandon Routh or Henry Cavill'') or Superman's alter ego ``Clark Kent''. But actually, Clark Kent is a mask and Kal-El is the real person, which humans refer to as Superman.\footnote{This holds according to the standard interpretation of Superman's identity, but there is a huge ongoing debate as to which of the identities (Superman/Kal-El or Clark Kent) is the real person and which is the fa\c{c}ade. Google ``Clark Kent'' and be prepared to be fascinated by the philosophical depth of the arguments. I recommend also watching the brilliant monologue on Superman that Bill delivers at the end of \emph{Kill Bill: Vol. 2}~\cite{KillBill2}.} Resolving or discovering the identity of an ``entity'' (a person or a digital entity such as a process, a client, or a server) is a challenging research problem, with applications in cybersecurity that range from discovering who really is the person behind an email address to attribution of cyber-attacks.

\begin{figure}[!t]
\centering
\includegraphics[width=0.8\textwidth]{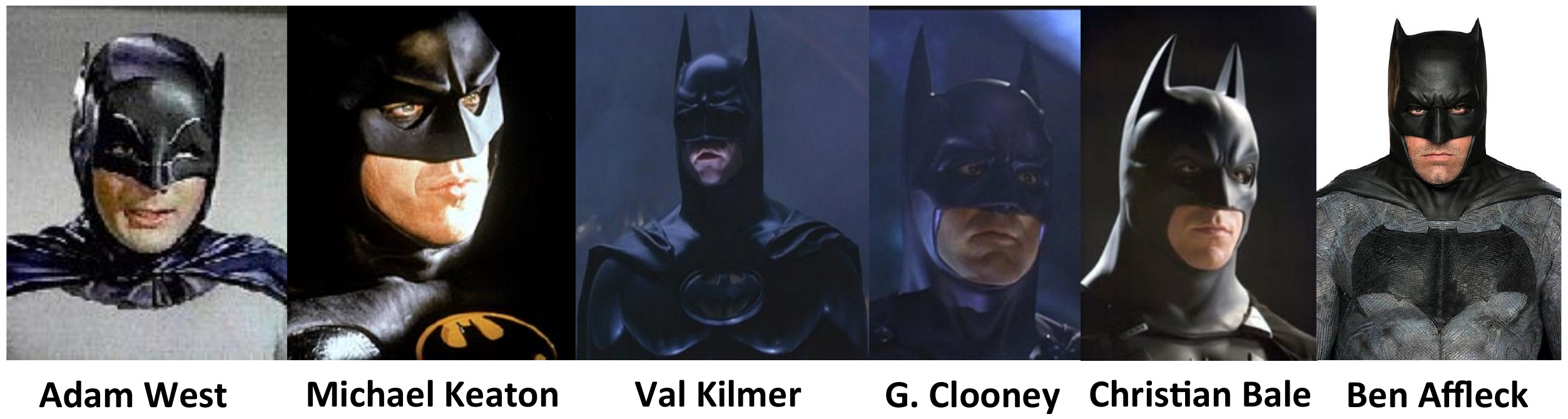}
\caption{Who is Batman?}
\label{fig:Batman}
\end{figure}


Several techniques and mechanisms have been designed to provide 
\emph{entity authentication}, i.e., let one party prove the identity of another party. Cryptographic protocols such as 
\emph{Kerberos} or the \emph{SAML 2.0 Web Browser Single Sign-On protocol} 
have been developed to that end. However, cryptographic protocols are notoriously difficult to get right and several attacks have been discovered by carrying out a formal analysis using automated tools, e.g., \cite{AVANTSSAR,Armando2008}. Films and artworks can help explain such attacks as well. For instance, the man-in-the-middle attack on the SAML-based Single Sign-On for Google Apps described in~\cite{Armando2008} is due to the certification authority simply (and wrongly) signing a certificate that says ``I certify that this is the client. Signed: The Certification Authority''... so that whoever is shown the certificate can first copy it and then claim to be the client.
This can be explained by analogy with the ``carte blanche'' issued to Milady De Winter by Cardinal Richelieu in \emph{The Three Musketeers} by Alexandre Dumas p\`ere (``It is by my order and for the benefit of the State that the bearer of this note has done what has been done. Signed: Richelieu'') and then used by d'Artagnan at the end of the novel to be pardoned for the execution of 
Milady.\footnote{One could similarly refer to the letters of transit in the movie \emph{Casablanca}~\cite{Casablanca}.}

\section{Concluding discussion: towards Explainable Security}


Such ``popular'' explanations are not meant to replace the mathematical definitions and explanations. On the contrary, the mathematical definitions that can be given of these security notions are a fundamental technical add on, on top of our intuition. The problem is that sometimes when teaching, or when explaining, cybersecurity, the experts forget (or don't know how) to refer to the intuition and only focus on the technical definition, thereby possibly frustrating and scaring off much of the audience. 
A clear and simple explanation, with something that they are already familiar with, such as a non-security related movie or novel like Spartacus or The Three Musketeers, would have made all audience members less irritated, stressed and annoyed, and thus more receptive. 
Explanations with popular films and the arts can reduce the mental effort required of the laypersons and increase their understanding, and ultimately their willingness to engage with security systems.

I have been using films to explain security for the last 15 years or so, both in my lectures and in public engagement talks, and this is the first of a series of papers that I plan to write on using popular films and artworks to explain cybersecurity. I will, in particular, consider other properties and security notions, and  investigate the links with 
\emph{usable security}, \emph{security awareness} and \emph{security economics} (see the references in~\cite{XSec}), and with the \emph{NASA-Task Load Index (NASA-TLX)} and other workload models.





It will also be interesting to investigate the relationships between explaining security with films and the arts and the growing body of literature on using film to explain and teach different disciplines such as (to name only a few):
\begin{itemize}
\item philosophy~\cite{Ariemma15,Ariemma17,Cabrera07,Mordacci15,Mordacci17},
\item history~\cite{Marcus18},
\item social sciences
\item management and organizational behavior~\cite{Champoux00a,Champoux00b},
\item mental illness~\cite{Rubin12}.
\end{itemize}
There is much to be learned from these approaches, and it might even be possible to adopt or adapt some of these techniques and tools to the case of explaining security. 
There is also the database recently developed by Blasco and Quaglia~\cite{BlascoQuaglia18} to use films for information security teaching: while their motivation and goal is quite different from the one of this paper, their database is very interesting and I plan to investigate possible synergies between their approach and mine.



%
%
%
\bibliographystyle{plain}
\bibliography{usingfilm}

\end{document}